\documentclass[aps,showpacs,prd,twocolumn]{revtex4}
\usepackage{amssymb}
\usepackage{amsfonts}
\usepackage{amsmath}
\usepackage{graphicx}
\usepackage{color}
\usepackage[dvips]{epsfig}
\usepackage[dvips]{graphicx}
\usepackage{float}
\usepackage{epstopdf}
\usepackage[colorlinks=true,pdfstartview=FitV,linkcolor=blue,citecolor=blue,urlcolor=blue,breaklinks=true]{hyperref}

\begin{document}

\title{Self-dual solitons in a $CPT$-odd and Lorentz-violating gauged $O(3)$
sigma model}
\author{R. Casana}
\email{rodolfo.casana@gmail.com}
\author{C. F. Farias}
\email{cffarias@gmail.com}
\author{M. M. Ferreira, Jr.}
\email{manojr.ufma@gmail.com}
\author{G. Lazar}
\email{gzsabito@gmail.com}
\affiliation{$^{1}$Departamento de F\'{\i}sica, Universidade Federal do Maranh\~{a}o,
65080-805, S\~{a}o Lu\'{\i}s, Maranh\~{a}o, Brazil.}

\begin{abstract}
We have performed a complete study of self-dual configurations in a $CPT$%
-odd and Lorentz-violating gauged $O(3)$ nonlinear sigma model. We have
consistently implemented the Bogomol'nyi-Prasad-Sommerfield (BPS) formalism and
obtained the correspondent differential first-order equations describing
electrically charged self-dual configurations. The total energy and
magnetic flux of the vortices, besides being proportional to the winding
number, also depend explicitly on the Lorentz-violating coefficients
belonging to the sigma sector. The total electrical charge is proportional
to the magnetic flux such as it occurs in Chern-Simons models. The
Lorentz violation in the sigma sector allows one to interpolate between
Lorentz-violating versions of some sigma models: the gauged $O(3)$ sigma
model and the Maxwell-Chern-Simons $O(3)$ sigma model. The Lorentz
violation enhances the amplitude of the magnetic field and BPS energy
density near the origin, augmenting the deviation in relation to the
solutions deprived of Lorentz violation.
\end{abstract}

\pacs{11.10.Lm,11.27.+d,12.60.-i, 74.25.Ha}
\maketitle

\section{Introduction}

In the early 1960s Gell-Mann and Levy \cite{Gell-Mann}, based on the works
of Schwinger \cite{Schwinger} and Polkinghorne \cite{Polkinghorne},
constructed a renormalizable field theory for the new scalar mesons $\sigma $
with null isotopic spin, being this theory called the sigma model. These
authors considered the possibility of modifying the sigma model by supposing
it as a composite of the pion field rather than an elementary one describing
a new particle. This was the arising of the nonlinear sigma model (NL$\sigma
$M), which featured an $O(4)$ symmetry initially. Later, the model
was endowed with an $O(3)$ symmetry, the $O(3)$ NL$\sigma $M \cite{polyakov},
started to gain special attention and was widely applied to study different
aspects of field theory and condensed matter physics\cite{CMP}. The $O(3)$ NL%
$\sigma $M is also interesting because it provides topological solitons
whose equations are exactly integrable in the Bogomol'nyi-Prasad-Sommerfield
(BPS) limit \cite{BPS}. The solutions of these BPS equations describe a map
from a spherical surface that represents the two-dimensional physical space
to a spherical surface in the internal field space, being classified
according to the second homotopy group $\Pi _{2}\left( S_{2}\right) =\mathbb{%
Z}$. This model, however, has a serious inconvenience because the solutions
are scale invariant preventing one from describing localized particles \cite%
{zakrzewski}. A way for breaking the scale invariance is gauging the $U(1)$
subgroup and providing a spontaneous symmetry breaking potential. Such
mechanism was suggested in Ref. \cite{schroers}, where a
Maxwell term controlling the gauge field dynamics was introduced, so that topological
solitons with arbitrary magnetic flux were engendered. A similar mechanism
was implemented in Ref. \cite{ghosh} with the Abelian Chern-Simons in the
gauge sector, implying topological and nontopological solitons. These two
approaches produce topological solitons which are infinitely degenerate in a
given topological sector. Such degenerescence was circumvented in Refs. \cite%
{mukherjee1,mukherjee2} by means of the introduction of a self-interacting
potential possessing a symmetry breaking minima. This potential induces a
new topology in which the infinite circle of physical space is mapped on the
equatorial circle of the internal space so that the solitons are now
classified by the first homotopy group $\Pi _{1}\left( S_{1}\right) =\mathbb{%
Z}$. The $O(3)$ sigma model has also been analyzed with
the gauge field dynamics ruled by both the Maxwell and Chern-Simons terms
\cite{sigmaMCSH1, sigmaMCSH2}. In the same context, but with a nonminimally
coupled gauge field, charged BPS soliton solutions were found \cite{almeida},
revealing a magnetic flux quantized only for topological solitons.

Investigation of topological defects in different theoretical contexts has
been a sensitive matter in the latest years. Among the new frameworks are
the $CPT$- and Lorentz-violating (LV) field theories. The violation of the $CPT$
and Lorentz symmetries is a theoretical possibility that has been
extensively investigated since 1996, mainly in the framework of the standard
model extension (SME) \cite{kost97,coleman}, with several repercussions \cite{LV1, LV2, LV3}.
In LV theories, the formation of defects was
considered in many situations, for example, in solitons generated by scalar
fields \cite{scalar}, general defects modified by tensor fields \cite%
{seifert}, Abelian monopoles \cite{barraz} and oscillons \cite{oscillons}.
BPS vortex solutions in Abelian Higgs models including Lorentz-violating
terms have been extensively examined in Refs. \cite%
{miller,casana1,sourrouille,belich,hott,Guillermo}. The effects of $CPT$%
-even LV terms on the topological BPS configurations of the gauged $O(3)$
sigma model were recently studied in Ref. \cite{Claudio}, with the
achievement of a generalization of the results of the gauged $O(3)$ sigma
models studied in Refs. \cite{mukherjee1,mukherjee2}.

In this manuscript, we analyze the self-dual structure of the gauged
$O(3)$ sigma model provided with $CPT$-odd and $CPT$-even
Lorentz-violating terms. More specifically, we have modified the dynamics
of the Maxwell sector with the $CPT$-odd Carroll-Field-Jackiw term
\cite{Jackiw}, whereas a $CPT$-even LV term was included in the $O(3)$ sigma
sector. The suitable projection of this Lagrangian in
(1+2)-dimensions yields a model similar to Maxwell-Chern-Simons gauged $O(3)$
sigma model (MCS$\sigma $M) including LV terms in
the scalar sigma sector. Next, we have implemented the BPS formalism and
written the self-dual equations describing the self-dual configurations of
this $CPT$-odd and LV model. To assess the repercussion of the LV terms on the
self-dual configurations of the gauged $O(3)$ sigma model, we have
found the rotationally symmetric vortex solutions. These solutions possess
finite energy proportional to the topological charge. The LV model provides
some interesting limits connecting different versions of the gauge $O(3)$%
 sigma model. The profiles are obtained numerically and compared
with the ones in the total absence of Lorentz-violation in order to highlight
the LV effects. Finally, we present our final remarks and conclusions.

\section{A $CPT$-odd and Lorentz-violating gauged $O(3)$ sigma model}

The (1+2)-dimensional gauged $O(3)$ sigma model (M$\sigma
$M) introduced in Refs. \cite{schroers,mukherjee1,mukherjee2} is
defined by the following Lagrangian density
\begin{equation}
\mathcal{L}_{M\sigma {M}}=-\frac{1}{4}F_{\mu \nu }F^{\mu \nu }+\frac{1}{2}%
D^{\mu }\vec{\phi}\cdot D_{\mu }\vec{\phi}-U(\vec{\phi}),  \label{msm}
\end{equation}%
where $A_{\mu }$ is the gauge field and $F_{\mu \nu
}=\partial _{\mu }A_{\nu }-\partial _{\nu }A_{\mu }$ is the Abelian
strength tensor field. The field $\vec{\phi}=\left( \phi
_{1},\phi _{2},\phi _{3}\right) $ is a triplet of real scalar
fields constituting a vector in the internal space, with fixed norm $\vec{%
\phi}\cdot \vec{\phi}=1$, describing the $O(3)$ NL$\sigma$M.
The coupling between the Abelian gauge and sigma field is ruled
by the minimal covariant derivative
\begin{equation}
D_{\mu }\vec{\phi}=\partial _{\mu }\vec{\phi}-A_{\mu }\hat{n}_{3}\times \vec{%
\phi},
\end{equation}%
with $\hat{n}_{3}$ being a unitary vector along the
3-direction in the internal scalar field space, while $U(\vec{\phi})$%
is the self-interacting potential.

In order to include Lorentz violation, we consider the
(1+3)-dimensional version of model (\ref{msm}) supplementing it with the $%
CPT$-odd Carroll-Field-Jackiw (CFJ) term in the gauge sector and a $%
CPT$-even term in the sigma field sector. This way, the $CPT$%
-odd Lagrangian density describing the proposed Lorentz-violating
sigma model is
\begin{eqnarray}
\mathcal{L} &=&-\frac{1}{4}F_{\mu \nu }F^{\mu \nu }-\frac{1}{4}\epsilon
^{\mu \nu \rho \sigma }\left( k_{AF}\right) _{\mu }A_{\nu }F_{\rho \sigma }
\label{Lag_CFJ} \\[0.15cm]
&&+\frac{1}{2}D^{\mu }\vec{\phi}\cdot D_{\mu }\vec{\phi}+\frac{1}{2}\left(
k_{\phi \phi }\right) ^{\mu \nu }D_{\mu }\vec{\phi}\cdot D_{\nu }\vec{\phi}%
-U.  \notag
\end{eqnarray}%
The four-vector $\left( k_{AF}\right) _{\alpha }$ is the
CFJ background with mass dimension $+1$. The dimensionless tensor $%
\left( k_{\phi \phi }\right) ^{\mu \nu }$ is real and symmetric,
containing the LV and $CPT$-even parameters in the sigma sector.
The potential $U$ describes some convenient interaction producing
self-dual configurations, still to be determined.

From the Lagrangian density (\ref{Lag_CFJ}), the equation of motion for the
gauge field reads
\begin{equation}
\partial _{\nu }F^{\nu \mu }+\frac{1}{2}\epsilon ^{\mu \alpha \rho \sigma
}\left( k_{AF}\right) _{\alpha }F_{\rho \sigma }=j^{\mu },  \label{ggeq}
\end{equation}%
where the conserved current density in this LV framework,
\begin{equation}
j^{\mu }=\left[ g^{\mu \nu }+{\left( k_{\phi \phi }\right) ^{\mu \nu }}%
\right] \hat{n}_{3}\cdot \left( \vec{\phi}\times D_{\nu }\vec{\phi}\right) ,
\end{equation}
is the counterpart of the one in absence of Lorentz violation, $j^{\mu }=%
\hat{n}_{3} \cdot \left( \vec{\phi}\times D^{\mu }\vec{\phi}\right)$,
displayed in Ref. \cite{mukherjee2}. The equation of motion for the sigma
field is
\begin{eqnarray}
\left[ g^{\mu \nu }+{\left( k_{\phi \phi }\right) ^{\mu \nu }}\right] D_{\mu
}D_{\nu }\vec{\phi} &=&\left( \vec{\phi}\cdot \frac{\partial U}{\partial
\vec{\phi}}\right) \vec{\phi}-\frac{\partial U}{\partial \vec{\phi}} \\%
[0.15cm]
&&\hspace{-1.5cm}+\left[ g^{\mu \nu }+{\left( k_{\phi \phi }\right) ^{\mu
\nu }}\right] \left( \vec{\phi}\cdot D_{\mu }D_{\nu }\vec{\phi}\right) \vec{%
\phi}.  \notag
\end{eqnarray}

From now on we are interested in the topological self-dual
configurations arising from the (1+2)-dimensional version of the model (%
\ref{Lag_CFJ}). For such a purpose to be fulfilled, we implement a
projection procedure doing $\partial _{3}\vec{\phi}=0$, $A_{3}=0$, $%
\partial _{3}A_{\mu }=0$ ($\mu =0,1,2$). Besides, it is
necessary to impose $(k_{\phi \phi })_{0i}=0$. In this way, from (%
\ref{ggeq}), the planar stationary Gauss law is
\begin{equation}
\partial _{j}\partial _{j}A_{0}-(k_{AF})_{3}B=[1+(k_{\phi \phi
})_{00}][(\phi _{1})^{2}+(\phi _{2})^{2}]A_{0},  \label{gauss}
\end{equation}%
and the planar stationary Ampere law reads
\begin{equation}
\epsilon _{ij}\partial _{j}B-(k_{AF})_{3}\epsilon _{ij}\partial
_{j}A_{0}=-[\delta _{ij}-(k_{\phi \phi })_{ij}]\hat{n}_{3}\!\cdot (\vec{\phi}%
\times \!D_{j}\vec{\phi}),  \label{ampere}
\end{equation}%
where $B\equiv B_{3}=F_{12}$ defines the magnitude of the magnetic field
along the $z$-axis and Latin indexes run over $i,j=1,2$. Here, we point out
that the magnetic field $B=F_{12}$ is effectively a planar or
"scalar" version of the 3D-magnetic field $B_{k}=\epsilon
_{kmn}F_{mn}/2$. Both equations clearly show that the LV
coefficient $(k_{AF})_{3}$ is responsible for the coupling between electric
and magnetic sectors, allowing, in principle, the occurrence of electrically
charged configurations. The $(k_{AF})_{3}$ coefficient in
both the Gauss and Ampere laws plays a similar role of the
Chern-Simons mass in the $(1+2)$-dimensional
Maxwell-Chern-Simons gauged $O(3)$ sigma model (MCS$\sigma $%
M) \cite{sigmaMCSH1,sigmaMCSH2}, defined by
\begin{eqnarray}
\mathcal{L}_{MCS\sigma {M}} &=&\mathcal{-}\frac{1}{4}F_{\alpha \beta
}F^{\alpha \beta }-\frac{1}{4}\kappa \epsilon ^{\beta \rho \sigma }A_{\beta
}F_{\rho \sigma }  \notag \\
&&+\frac{1}{2}D^{\alpha }\vec{\phi}\cdot D_{\alpha }\vec{\phi}+\frac{1}{2}%
\partial _{\mu }\Psi \partial ^{\mu }\Psi  \label{LSC} \\
&&-\frac{1}{2}[(\phi _{1})^{2}+(\phi _{2})^{2}]\Psi ^{2}-U\left( \left\vert
\phi \right\vert ,\Psi \right),  \notag
\end{eqnarray}%
where $\kappa $ is the Chern-Simons mass, $U\left(
\left\vert \phi \right\vert ,\Psi \right) $ is the self-dual
potential and $\Psi $ is a neutral field. The introduction of this
neutral scalar field is a well established procedure for a consistent
description of self-dual configurations, and it was first reported in the
context of Maxwell-Chern-Simons-Higgs models \cite{lee} based in
supersymmetric arguments. It was also successfully implemented in other
subsequent extensions \cite{bolog}, including to describe charged
topological configurations in Lorentz-violating models \cite%
{casana1,Guillermo,Claudio}. This Lagrangian density was initially
proposed (but not solved) in Ref. \cite{sigmaMCSH1}, once its focus was in
the Chern-Simons O(3) gauge sigma model. It was also addressed in
Ref. \cite{sigmaMCSH2}, where the self-duality of the MCS$\sigma $M
was considered, but without developing the numerical solutions for the
self-dual configurations. Here, the numerical solutions for MCS$%
\sigma $ M will be achieved, solved and used as a suitable background
for physical comparisons.

A consistent description of the self-dual configurations carrying electric
field in (1+3) dimensions also requires the introduction of the
neutral scalar field appearing Lagrangian (\ref{LSC}), being given by the
following (1+3)-dimensional model:
\begin{eqnarray}
\mathcal{L} &=&\mathcal{-}\frac{1}{4}F_{\alpha \beta }F^{\alpha \beta }-%
\frac{1}{4}\epsilon ^{\alpha \beta \rho \sigma }\left( k_{AF}\right)
_{\alpha }A_{\beta }F_{\rho \sigma }  \notag \\[0.15cm]
&&+\frac{1}{2}D^{\alpha }\vec{\phi}\cdot D_{\alpha }\vec{\phi}+\frac{1}{2}%
\left( k_{\phi \phi }\right) ^{\alpha \beta }D_{\alpha }\vec{\phi}\cdot
D_{\beta }\vec{\phi}  \notag \\[0.15cm]
&&+\frac{1}{2}\partial _{\mu }\Psi \partial ^{\mu }\Psi -\frac{1}{2}%
[1+(k_{\phi \phi })_{00}][(\phi _{1})^{2}+(\phi _{2})^{2}]\Psi ^{2}  \notag
\\[0.15cm]
&&-U\left( \left\vert \phi \right\vert ,\Psi \right) ,  \label{L4}
\end{eqnarray}%
where $U\left( \left\vert \phi \right\vert ,\Psi \right) $ is a convenient
potential providing charged BPS configurations. The
(1+2)-dimensional projection of the Lagrangian density (\ref{L4}) engenders
a modified version of the MCS$\sigma $M model (\ref{LSC}) with the
CFJ term becoming $\left( k_{AF}\right) _{3}\epsilon ^{\beta \rho \sigma
}A_{\beta }F_{\rho \sigma }$, and the coefficient $(k_{AF})_{3}$
playing the role of the Chern-Simons mass $\left( \kappa \right) $%
. The difference rests in the presence of LV terms in the sigma
sector.

\section{The BPS formalism}

We are interested in finding first-order self-dual differential (BPS)
equations whose solutions are minimum energy configurations that also solve
the second-order Euler-Lagrange equations. To carry out the BPS procedure,
we first write the stationary energy of the (1+2)-dimensional version of the
model (\ref{L4}),
\begin{eqnarray}
E &=&\int d^{2}x\left\{ \frac{1}{2}\left[ \delta _{ij}-\left( k_{\phi \phi
}\right) _{ij}\right] D_{i}\vec{\phi}\cdot D_{j}\vec{\phi}\right.  \notag \\%
[0.15cm]
&&+\frac{1}{2}B^{2}+U+\frac{1}{2}\left( \partial _{j}A_{0}\right) ^{2}+\frac{%
1}{2}\left( \partial _{j}\Psi \right) ^{2}  \notag \\[0.15cm]
&&+\frac{1}{2}\left[ 1+\left( k_{\phi \phi }\right) _{00}\right] \left[
\left( \phi _{1}\right) ^{2}+\left( \phi _{2}\right) ^{2}\right] \left(
A_{0}\right) ^{2}  \notag \\[0.15cm]
&&\left. +\frac{1}{2}\left[ 1+{\left( k_{\phi \phi }\right) }_{00}\right] %
\left[ \left( \phi _{1}\right) ^{2}+\left( \phi _{2}\right) ^{2}\right] \Psi
^{2}\right\} ,  \label{ED1}
\end{eqnarray}%
which is positive definite as long as
\begin{equation}
\delta _{jk}-\left( k_{\phi \phi }\right) _{jk}>0, \left( k_{\phi
\phi }\right) _{00}>-1.
\end{equation}

In order to implement the BPS procedure, we define $\tilde{D}_{k}\vec{\phi}%
=M_{kj}D_{j}\vec{\phi},$ which allows one to write
\begin{eqnarray}
\left[ \delta _{ij}-\left( k_{\phi \phi }\right) _{ij}\right] D_{i}\vec{\phi}%
\cdot D_{j}\vec{\phi} &=&\tilde{D}_{k}\vec{\phi}\cdot \tilde{D}_{k}\vec{\phi}
\\
&=&M_{ki}M_{kj}D_{i}\vec{\phi}\cdot D_{j}\vec{\phi},  \notag
\end{eqnarray}%
where the $M_{ij}$ are the elements of the matrix $\mathbb{M}$ englobing the
spatial LV parameters of the sigma sector, being defined as
\begin{equation}
M_{ki}M_{kj}=\delta _{ij}-{\left( k_{\phi \phi }\right) _{ij}}.
\end{equation}

By introducing the identity,
\begin{eqnarray}
\frac{1}{2}\tilde{D}_{k}\vec{\phi}\cdot \tilde{D}_{k}\vec{\phi} &=&\frac{1}{4%
}\left( \tilde{D}_{j}\vec{\phi}\pm \epsilon _{jm}\vec{\phi}\times \tilde{D}%
_{m}\vec{\phi}\right) ^{2}  \label{trick} \\[0.15cm]
&&\mp \left( \det \mathbb{M}\right) \phi _{3}B\pm \left( \det \mathbb{M}%
\right) \epsilon _{ik}\partial _{i}\left( A_{k}\phi _{3}\right)  \notag \\%
[0.15cm]
&&\pm \left( \det \mathbb{M}\right) \vec{\phi}\cdot \left( \partial _{1}\vec{%
\phi}\times \partial _{2}\vec{\phi}\right) ,  \notag
\end{eqnarray}%
the energy (\ref{ED1}) is expressed as
\begin{eqnarray}
E &=&\int d^{2}x\left\{ \frac{1}{4}\left( \tilde{D}_{j}\vec{\phi}\pm
\epsilon _{jm}\vec{\phi}\times \tilde{D}_{m}\vec{\phi}\right) ^{2}\right.
\notag \\
&&+\frac{1}{2}\left( B\mp \sqrt{2U}\right) ^{2}+\frac{1}{2}\left( \partial
_{j}A_{0}\pm \partial _{j}\Psi \right) ^{2}  \notag \\
&&\frac{1}{2}\left[ 1+\left( k_{\phi \phi }\right) _{00}\right] \left[
\left( \phi _{1}\right) ^{2}+\left( \phi _{2}\right) ^{2}\right] \left[
A_{0}\pm \Psi \right] ^{2}  \notag \\
&&\pm \left( \det \mathbb{M}\right) \left[ \vec{\phi}\cdot \left( \partial
_{1}\vec{\phi}\times \partial _{2}\vec{\phi}\right) +\epsilon _{ik}\partial
_{i}\left( A_{k}\phi _{3}\right) \right]  \notag \\
&&\pm B\sqrt{2U}\mp \left( \det \mathbb{M}\right) \phi _{3}B\mp \left(
\partial _{j}\Psi \right) \left( \partial _{j}A_{0}\right)  \notag \\
&&\left. \frac{{}}{{}}\mp \left[ 1+\left( k_{\phi \phi }\right) _{00}\right] %
\left[ \left( \phi _{1}\right) ^{2}+\left( \phi _{2}\right) ^{2}\right]
A_{0}\Psi \right\} .  \label{ED2}
\end{eqnarray}%
Using the Gauss law (\ref{gauss}), the last row reads as
\begin{equation}
\mp \Psi \partial _{j}\partial _{j}A_{0}\pm \left( k_{AF}\right) _{3}B\Psi,
\end{equation}
so that the energy reads
\begin{eqnarray}
E &=&\int d^{2}x\left\{ \frac{1}{4}\left( \tilde{D}_{j}\vec{\phi}\pm
\epsilon _{jm}\vec{\phi}\times \tilde{D}_{m}\vec{\phi}\right) ^{2}\right.
\notag \\
&&\hspace{-0.5cm}+\frac{1}{2}\left( B\mp \sqrt{2U}\right) ^{2}+\frac{1}{2}%
\left( \partial _{j}A_{0}\pm \partial _{j}\Psi \right) ^{2}  \notag \\
&&\hspace{-0.5cm}+\frac{1}{2}\left[ 1+\left( k_{\phi \phi }\right) _{00}%
\right] \left[ \left( \phi _{1}\right) ^{2}+\left( \phi _{2}\right) ^{2}%
\right] \left[ A_{0}\pm \Psi \right] ^{2}  \label{ED3} \\
&&\hspace{-0.5cm}\pm \left( \det \mathbb{M}\right) \left[ \vec{\phi}\cdot
\left( \partial _{1}\vec{\phi}\times \partial _{2}\vec{\phi}\right)
+\epsilon _{ik}\partial _{i}\left( A_{k}\phi _{3}\right) \right]  \notag \\
&&\hspace{-0.5cm}\left. \!\!\pm B\left[ \sqrt{2U}-\left( \det \mathbb{M}%
\right) \phi _{3}+\left( k_{AF}\right) _{3}\Psi \right] \mp \partial
_{j}\left( \Psi \partial _{j}A_{0}\right) \right\} .  \notag
\end{eqnarray}%
In the fifth row, one requires the factor multiplying the magnetic field to
be null, which leads to the BPS potential
\begin{equation}
U=\frac{1}{2}\left[ \left( \det \mathbb{M}\right) \phi _{3}-\left(
k_{AF}\right) _{3}\Psi \right] ^{2}.  \label{PM}
\end{equation}

The integration of the expression in the fourth row of Eq. (\ref{ED3}),
\begin{equation}
T_{0}=\frac{\left( \det \mathbb{M}\right) }{4\pi }\!\int \!d^{2}x\!\left[
\vec{\phi}\cdot \left( \partial _{1}\vec{\phi}\times \partial _{2}\vec{\phi}%
\right) +\epsilon _{ik}\partial _{i}\left( A_{k}\phi _{3}\right) \right] ,
\label{tpc}
\end{equation}%
provides the topological charge of the model, which shows dependence on the
Lorentz-violating coefficients belonging to the sigma sector \cite{Claudio}.
As it was also reported in Ref. \cite{Claudio}, the topological conserved
current is
\begin{equation}
K_{\mu }=\frac{\left( \det \mathbb{M}\right) }{8\pi }\epsilon _{\mu \alpha
\beta }\left[ \vec{\phi}\cdot \left( D^{\alpha }\vec{\phi}\times D^{\beta }%
\vec{\phi}\right) +F^{\alpha \beta }\phi _{3}\right] ,
\end{equation}%
whose component $K_{0},$\ whenever integrated over the space, yields the
conserved topological charge (\ref{tpc}). By considering the fields $\Psi $
and $A_{0}$ going to zero at infinity, in the fifth row of Eq. (\ref{ED3})
the integration of the term $\partial _{j}\left( \Psi \partial
_{j}A_{0}\right) $ gives null contribution to the energy. Thus, the energy
of the solutions becomes%
\begin{eqnarray}
E &=&4\pi T_{0}  \label{ED4} \\
&&+\int d^{2}x\left\{ \frac{1}{2}\left( B\mp \left[ \left( \det \mathbb{M}%
\right) \phi _{3}-\left( k_{AF}\right) _{3}\Psi \right] \right) ^{2}\right.
\notag \\
&&+\frac{1}{4}\left( \tilde{D}_{j}\vec{\phi}\pm \epsilon _{jm}\vec{\phi}%
\times \tilde{D}_{m}\vec{\phi}\right) ^{2}+\frac{1}{2}\left( \partial
_{j}A_{0}\pm \partial _{j}\Psi \right) ^{2}  \notag \\
&&\left. +\frac{1}{2}\left[ 1+\left( k_{\phi \phi }\right) _{00}\right] %
\left[ \left( \phi _{1}\right) ^{2}+\left( \phi _{2}\right) ^{2}\right] %
\left[ A_{0}\pm \Psi \right] ^{2}\right\} .  \notag
\end{eqnarray}%
This equation allows us to establish that the energy has a lower bound given
by
\begin{equation}
E\geq \pm 4\pi T_{0},
\end{equation}
attained whenever the fields satisfy the following self-dual or BPS
equations,
\begin{equation}
\tilde{D}_{j}\vec{\phi}\pm \epsilon _{jm}\vec{\phi}\times \tilde{D}_{m}\vec{%
\phi}=0,
\end{equation}%
\begin{equation}
B=\pm \left[ \left( \det \mathbb{M}\right) \phi _{3}-\left( k_{AF}\right)
_{3}\Psi \right] ,
\end{equation}%
\begin{equation}
\partial _{i}A_{0}\pm \partial _{i}\Psi =0,
\end{equation}%
\begin{equation}
A_{0}\pm \Psi =0.
\end{equation}

The condition $\Psi =\mp A_{0}$ saturates the two last equations and the
self-dual charged configurations are described by
\begin{equation}
\tilde{D}_{j}\vec{\phi}\pm \epsilon _{jm}\vec{\phi}\times \tilde{D}_{m}\vec{%
\phi}=0,  \label{bq1}
\end{equation}%
\begin{equation}
B=\pm \left( \det \mathbb{M}\right) \phi _{3}+\left( k_{AF}\right) _{3}A_{0},
\label{bq2}
\end{equation}%
with the modified Gauss law
\begin{equation}
\partial _{j}\partial _{j}A_{0}-(k_{AF})_{3}B=[1+{(k_{\phi \phi })}%
_{00}][(\phi _{1})^{2}+(\phi _{2})^{2}]A_{0}.  \label{bq3}
\end{equation}%
It is clear that, for null Lorentz-violating parameters, we recover the BPS
equations of the gauged $O(3)$ sigma model (\ref{msm}). On the other hand,
if we set to be null only the Lorentz-violating parameters of the sigma
sector we recuperate the self-dual equations of the Maxwell-Chern-Simons $%
O(3)$ sigma model (\ref{LSC}) with $(k_{AF})_{3}=\kappa$. Moreover, the
Gauss law (\ref{bq3}) implies the proportionality relation,
\begin{equation}
Q=\frac{(k_{AF})_{3}}{1+{(k_{\phi \phi })}_{00}}\Phi ,  \label{Rel}
\end{equation}%
between the total charge $(Q)$ of the self-dual configurations and the total
magnetic flux $(\Phi )$,
\begin{eqnarray}
Q &=&-\int d^{2}x~[(\phi _{1})^{2}+(\phi _{2})^{2}]A_{0},  \label{carga} \\
\Phi &=&\int d^{2}x~B.  \label{fluxo}
\end{eqnarray}

Our analysis will also compare the Lorentz-violating solutions with
the profiles of the corresponding models without Lorentz violation at all,
provided by the Lagrangian densities (\ref{msm}) and (\ref{LSC}). The
self-dual equation for the gauge $O(3)$ sigma model (\ref{msm}) are given by

\begin{equation}
D_{j}\vec{\phi}\pm \epsilon _{jm}\vec{\phi}\times D_{m}\vec{\phi}=0,
\label{bps1_msm}
\end{equation}%
\begin{equation}
B=\pm \phi _{3},  \label{bps2_msm}
\end{equation}%
whereas for the MCS$\sigma $M model (\ref{LSC}), the BPS
equations and Gauss law describing self-dual configurations read as
\begin{equation}
D_{j}\vec{\phi}\pm \epsilon _{jm}\vec{\phi}\times D_{m}\vec{\phi}=0,
\label{bps1_mcssm}
\end{equation}%
\begin{equation}
B=\pm \phi _{3}+\kappa A_{0},  \label{bps2_mcssm}
\end{equation}%
\begin{equation}
\partial _{j}\partial _{j}A_{0}-\kappa B=[(\phi _{1})^{2}+(\phi
_{2})^{2}]A_{0},  \label{bps3_mcssm}
\end{equation}%
respectively. For this latter case, the charge and magnetic flux
fulfill, $Q=\kappa \Phi $.

In the sequel we study the axially symmetrical self-dual solutions
describing electrically charged vortices in $CPT$-odd
Lorentz-violating framework, comparing them with solutions of the usual
models preserving Lorentz invariance.

\section{Axially symmetrical self-dual charged vortices}

For the energy to be finite, the field $\vec{\phi}$ should go asymptotically
to one of the minimum configurations of the potential, stated in Eq. (\ref%
{PM}). This is reached following the \textit{Ansatz} introduced in Ref. \cite%
{Claudio} for axially symmetric vortices in the presence of Lorentz violation,
\begin{eqnarray}
\phi _{1} &=&\sin g(r)\cos \left( \frac{n}{\Lambda }\theta \right) ,~\phi
_{2}=\sin g(r)\sin \left( \frac{n}{\Lambda }\theta \right) ,  \notag \\%
[-0.15cm]
&&  \label{ansatz} \\[-0.15cm]
\phi _{3} &=&\cos g(r),~A_{\theta }=-\frac{1}{r}\left[ a(r)-\frac{n}{\Lambda
}\right] ,~A_{0}=A_{0}(r),  \notag
\end{eqnarray}%
with the radial functions, $g(r)$, $a(r)$ and $A_0(r)$ being well behaved
and satisfying the following boundary conditions (see Sec. \ref{BC-cfj}):
\begin{eqnarray}
g(0) &=&0,~\ a(0)=\frac{n}{\Lambda },~\ A_{0}^{\prime }\left( 0\right) =0,
\notag \\[-0.15cm]
&&  \label{bcx} \\[-0.15cm]
g(\infty ) &=&\frac{\pi }{2},~\ a(\infty )=0,~\ A_{0}\left( \infty \right)
=0,  \notag
\end{eqnarray}%
which are compatible with the vacuum configurations of the potential for $%
r\rightarrow \infty$, while providing consistent solutions at $r=0$. The
non-null integer $n$ is the winding number of the self-dual vortices. The
constant $\Lambda$ is defined in terms of the Lorentz-violating parameters
belonging to the sigma sector,
\begin{equation}
\Lambda =\sqrt{\frac{1-\left( k_{\phi \phi }\right) _{\theta \theta }}{%
1-\left( k_{\phi \phi }\right) _{rr}}}.  \label{Lambda}
\end{equation}

In the \textit{Ansatz} (\ref{ansatz}), the magnetic field $B\ $reads%
\begin{equation}
B(r)=-\frac{a^{\prime }}{r},
\end{equation}%
where $(^{\prime })$ stands for the radial derivative. The BPS equations (%
\ref{bq1}) and (\ref{bq2}), projected on the \emph{Ansatz} (\ref{ansatz}),
become
\begin{eqnarray}
g^{\prime }&=&\pm \Lambda \frac{a}{r}\sin g,  \label{eqt2a} \\
-\frac{a^{\prime }}{r}&=&\pm \eta \cos g+(k_{AF})_{3}A_{0},  \label{eqt2b}
\end{eqnarray}%
whereas the Gauss law (\ref{bq3}) reads%
\begin{equation}
A_{0}^{\prime \prime }+\frac{A_{0}^{\prime }}{r}-(k_{AF})_{3}B=\eta \Lambda
\Delta A_{0}\sin ^{2}g,  \label{eqt2c}
\end{equation}%
with the parameters $\Delta $ and $\eta $ given by
\begin{eqnarray}
\Delta &=&\frac{1+{(k_{\phi \phi })}_{00}}{\eta \Lambda },  \label{Delta} \\
\eta &=&\det \mathbb{M=}\sqrt{\left[ 1-\left( k_{\phi \phi }\right) _{\theta
\theta }\right] \left[ 1-\left( k_{\phi \phi }\right) _{rr}\right] }.
\label{eta}
\end{eqnarray}

We use the BPS equations and the Gauss law to express the BPS energy density
as
\begin{equation}
\mathcal{E}_{BPS}=B^{2}+\eta \Lambda \left( \frac{a}{r}\sin g\right)
^{2}+\eta \Lambda \Delta \left( A_{0}\sin g\right) ^{2}+\left( A_{0}^{\prime
}\right) ^{2},  \label{defen}
\end{equation}%
which is positive definite because $\eta ,\Lambda ,\Delta >0.$

Replacing the \emph{Ansatz} (\ref{ansatz}) and the boundary conditions (\ref%
{bcx}) in Eq. (\ref{tpc}), the resulting topological charge is
\begin{equation}
T_{0}=\frac{n}{2}\frac{\eta }{\Lambda }.  \label{tpch}
\end{equation}

Moreover, under boundary conditions (\ref{bcx}), the magnetic flux (\ref%
{fluxo}) and the electric charge (\ref{carga}) become
\begin{eqnarray}
\Phi &=&2\pi \frac{n}{\Lambda },  \label{fluxm} \\[0.15cm]
Q &=&2\pi \frac{(k_{AF})_{3}}{\eta \Lambda \Delta }\frac{n}{\Lambda },
\label{charge_cfj}
\end{eqnarray}%
being both proportional to the winding number, $n$.

The topological charge (\ref{tpch}) and the magnetic flux (\ref%
{fluxm}) differ from the Lorentz symmetric ones, $T_{0}=n/2$ and $%
\Phi =2\pi n$, by the LV factors $\eta ,\Lambda $. The
self-dual vortices of the Lorentz-symmetric MCS$\sigma $M model are
described by
\begin{eqnarray}
g^{\prime } &=&\pm \frac{a}{r}\sin g, \\
-\frac{a^{\prime }}{r} &=&\pm \cos g+{\kappa }A_{0}, \\
A_{0}^{\prime \prime }+\frac{A_{0}^{\prime }}{r}-{\kappa }B &=&A_{0}\sin
^{2}g,
\end{eqnarray}%
while the correspondent BPS energy density is
\begin{equation}
\mathcal{E}_{BPS}=B^{2}+\left( \frac{a}{r}\sin g\right) ^{2}+\left(
A_{0}\sin g\right) ^{2}+\left( A_{0}^{\prime }\right) ^{2}.
\end{equation}

We also write self-dual the equations for the neutral vortices of
the Lorentz-symmetric gauged $O(3)$ sigma model, %
\begin{eqnarray}
g^{\prime } &=&\pm \frac{a}{r}\sin g, \\
-\frac{a^{\prime }}{r} &=&\pm \cos g,
\end{eqnarray}%
whose BPS energy density is
\begin{equation}
\mathcal{E}_{BPS}=B^{2}+\left( \frac{a}{r}\sin g\right) ^{2}.
\end{equation}

\subsection{Behavior of the profiles at boundaries\label{BC-cfj}}

We study the behavior of the solutions at boundaries by solving the BPS
equations and the Gauss law (\ref{eqt2c}) at the limits $r\rightarrow 0$ and
$r\rightarrow \infty $. Close to the origin, we obtain the following
expansions
\begin{eqnarray}  \label{A0}
g(r) &\approx &G_{n}r^{n}+\cdots,  \label{ggg2} \\[0.08in]
a(r) &\approx &\frac{n}{\Lambda }-\frac{[ev^{2}+(k_{AF})_{3}A_{0}(0)]}{2}%
er^{2}+\cdots,  \label{BH_a0} \\[0.08in]
A_{0}(r) &\approx &A_{0}(0)+\frac{[ev^{2}+(k_{AF})_{3}A_{0}(0)]}{4}%
(k_{AF})_{3}r^{2}+\cdots  \notag \\
&&  \label{A0_2}
\end{eqnarray}%
We observe that Eq. (\ref{BH_a0}) justifies the use of the modified \textit{%
Ansatz} (\ref{bcx}) and the boundary condition for $a(0)$. The constant $%
A_{0}(0)$ in (\ref{A0_2}) is determined numerically for every $n$. Moreover,
from (\ref{A0_2}) it is clear that the electric field must be null at
origin, i.e., $A^{\prime }(0)=0$, as stated in Eq. (\ref{bcx}).

At $r\rightarrow \infty $, the profiles present the following asymptotic
behavior:
\begin{eqnarray}
g(r) &\approx &\frac{\pi }{2}-C_{_{\infty }}\frac{e^{-mr}}{\sqrt{r}}+\cdots~,~~
\\[0.06in]
a(r) &\approx &\frac{mC_{_{\infty }}}{\Lambda }\sqrt{r}e^{-mr}+\cdots~,~\ \  \\
A_{0}(r) &\approx &\frac{C_{\infty }\left( m^{2}-\eta \Lambda \right) }{%
\Lambda \left( k_{AF}\right) _{3}}\frac{e^{-mr}}{\sqrt{r}}+\cdots,
\label{bc-3a}
\end{eqnarray}%
where $C_{_{\infty }}$ is a positive constant determined numerically. Thus,
the profiles behave in a similar way to the vortices of
Abrikosov-Nielsen-Olesen \cite{ANO}. The parameter $m$ is real and positive,
\begin{eqnarray}
m &=&\frac{1}{2}\sqrt{\left( k_{AF}\right) _{3}^{2}+\eta \,\Lambda \left( 1+%
\sqrt{\Delta }\right) ^{2}}  \label{betax} \\
&&-\frac{1}{2}\sqrt{\left( k_{AF}\right) _{3}^{2}+\eta \,\Lambda \left( 1-%
\sqrt{\Delta }\right) ^{2}},  \notag
\end{eqnarray}%
which is associated to the mass of the self-dual bosons and to the extension
of the defect.

Now we analyze the behavior of the profiles by studying some values of the
Lorentz-violating parameters involved in the bosonic mass. For fixed $\left(
k_{AF}\right) _{3}$, the behavior of the vortex profiles is governed by
the LV coefficients belonging only to the sigma sector. In this
scenario we can analyze two situations of interest: the first one occurs
when $\Delta =1$, providing
\begin{equation}
m=\frac{1}{2}\sqrt{\left( k_{AF}\right) _{3}^{2}+4\eta \,\Lambda }-\frac{1}{2%
}\left\vert \left( k_{AF}\right) _{3}\right\vert .  \label{m1}
\end{equation}%
This is the mass the MCS$\sigma $M \cite{sigmaMCSH1},\cite%
{sigmaMCSH2} would have in the presence of Lorentz violation only in the
sigma sector. Indeed, for $\eta =\Lambda =1$ (absence of LV terms
in the sigma sector), it becomes
\begin{equation}
m=\frac{1}{2}\sqrt{\left( k_{AF}\right) _{3}^{2}+4}-\frac{1}{2}\left\vert
\left( k_{AF}\right) _{3}\right\vert ,
\end{equation}%
the same mass of the Maxwell-Chern-Simons $O(3)$ sigma model \cite%
{sigmaMCSH2}.

The second regime to be highlighted happens when $\Delta $ takes
sufficiently large values [$\Delta \gg \left( k_{AF}\right) _{3}$]:%
\begin{equation}
m\rightarrow \sqrt{\eta \Lambda },  \label{m2}
\end{equation}%
corresponding to the bosonic mass of the gauged $O(3)$ sigma model
of Ref. \cite{Claudio} with Lorentz violation in the sigma sector (only).

Another interesting limit occurs by fixing the Lorentz-violating parameters
of the sigma sector and considering sufficiently large values of $\left(
k_{AF}\right) _{3}$, that is
\begin{equation}
m\rightarrow \frac{\eta \Lambda \sqrt{\Delta }}{\left( k_{AF}\right) _{3}},
\label{m3}
\end{equation}%
which corresponds to the mass the self-dual vortices of the Chern-Simons $%
O(3)$ sigma model of Ref. \cite{ghosh},\cite{sigmaMCSH1} would possess
considering Lorentz violation only in the sigma sector. In this
limit, taking $\left( k_{\phi \phi }\right) _{\mu \nu }=0$ or $%
\eta =\Lambda =\Delta =1$ leads to the Lorentz symmetric mass $%
(1/\left( k_{AF}\right) _{3}).$

\subsection{Numerical analysis}

Below, we depict the profiles obtained from numerical solutions of Eqs. (\ref%
{eqt2a})-(\ref{eqt2c}), under boundary conditions (\ref{bcx}), for winding
number $n=1$. We have fixed the Lorentz violating parameters $\Lambda =1.25$%
, $\eta =2$ and $\left( k_{AF}\right) _{3}=1.5$, allowing the parameter $%
\Delta $ to be free. Because the BPS energy density (\ref{defen}) is
positive definite for $\Delta >0$, we consider two regions: $0<\Delta <1$
(blue lines) and $\Delta >1$ (orange lines), in which the behavior of the
solutions are different.

There are two interesting values or limits of $\Delta $ that allow one to
recover the behavior of two known models in the presence of LV
coefficients. The first one is $\Delta =1$ (solid magenta
line), whose profiles correspond to Maxwell-Chern-Simons $O(3)$
sigma model with Lorentz violation only in the sigma sector [see comment
after Eq. (\ref{m1})]. The second one is the limit $\Delta \rightarrow
\infty $ (solid black line), yielding the gauged $O(3)$
sigma model with Lorentz violation only in the sigma sector \cite{Claudio}
[see comment after Eq. (\ref{m2})]. We also have depicted the symmetric
profiles corresponding to: (i) the Maxwell-Chern-Simons gauged $O(3)$%
 sigma model (\ref{LSC}) (green solid lines) with $\kappa =\left(
k_{AF}\right) _{3}=1.5$ and $\left( k_{\phi \phi }\right) _{\mu \nu }=0$%
, (ii) the gauged $O(3)$ sigma model (\ref{msm}) (red
solid lines) which corresponds to the total absence of Lorentz-violation [$%
\left( k_{AF}\right) _{\mu }=0,\;\left( k_{\phi \phi }\right) _{\mu \nu }=0$%
]. These two cases will serve as comparation basis to the
Lorentz-violating profiles.

\begin{figure}[]
\centering\includegraphics[width=8.5cm]{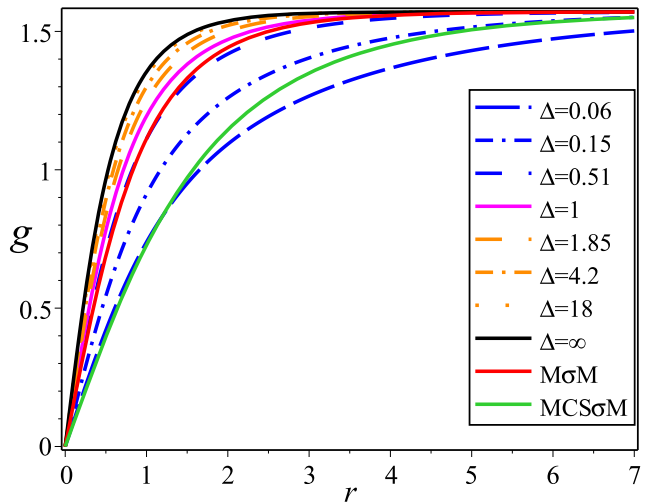}
\caption{Sigma field $g(r)$ profiles.}
\label{S_BPSx}
\end{figure}

Figure \ref{S_BPSx} depicts the profiles of the sigma field. For $0<\Delta
<1 $, the profiles are more spread and saturate the asymptotic value $\pi /2$
more slowly when $\Delta \rightarrow 0$ (blue lines). On the other hand, for
$\Delta >1$, the profiles are progressively narrower for growing $\Delta $
(orange lines), the maximum tightness being achieved in the limit $%
\Delta\rightarrow \infty $ (solid black line). Thus, the profiles for $%
\Delta >1$ are confined between the models described by $\Delta =1$ and $%
\Delta\rightarrow \infty $.

\begin{figure}[]
\centering\includegraphics[width=8.5cm]{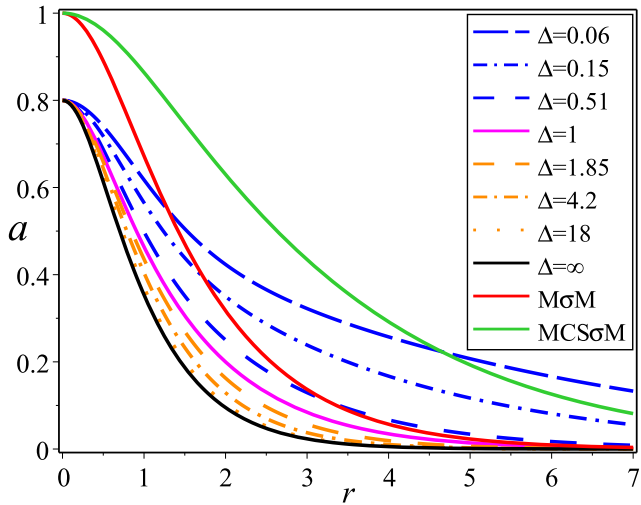}
\caption{Vector potential ${a}(r)$ profiles. }
\label{A_BPS}
\end{figure}

A similar description can be given for the profiles of the vector field $%
a(r) $, presented in Fig. \ref{A_BPS}. It is worthwhile to note that the
gauge field value at the origin changes from $a(0)=n$ (in the absence of LV) to $%
a(0)=n/\Lambda $ (in the presence of LV in the sigma sector), which is
evident in this graphic.

\begin{figure}[]
\centering
\includegraphics[width=8.5cm]{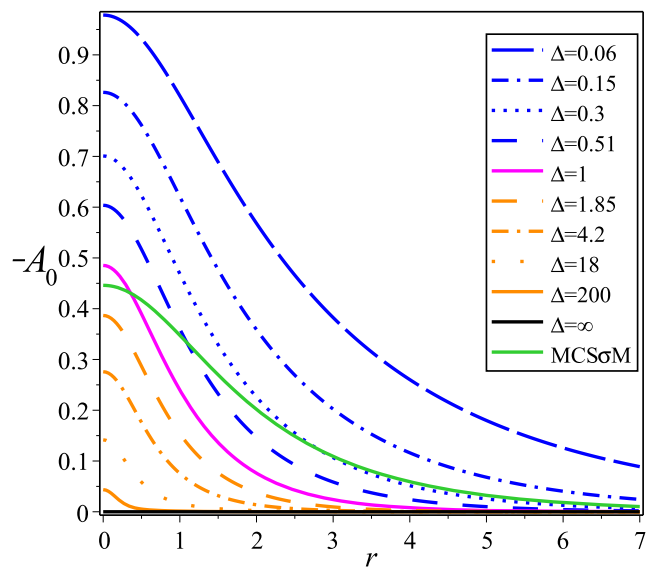}
\caption{Scalar potential $A_0 (r)$ profiles. }
\label{W_BPS}
\end{figure}

Figure \ref{W_BPS} depicts the profiles for the scalar potential. For $%
0<\Delta <1$ (blue lines), the profiles are more extended and with greater
intensity at the origin. The influence of the CFJ parameter, $(k_{AF})_{3}$,
is more pronounced when $\Delta \rightarrow 0$, while for $\Delta >1$
(orange lines) its effect becomes negligible for increasing values of $%
\Delta $. So, for large values of $\Delta $ the profiles become smaller and
smaller, overlapping the horizontal axis in the limit $\Delta \rightarrow
\infty $ (solid black line). It means that the vortices become electrically
neutral such as the ones of the usual Lorentz-invariant gauged $O(3)$ sigma
model, in this limit.

\begin{figure}[]
\centering
\includegraphics[width=8.5cm]{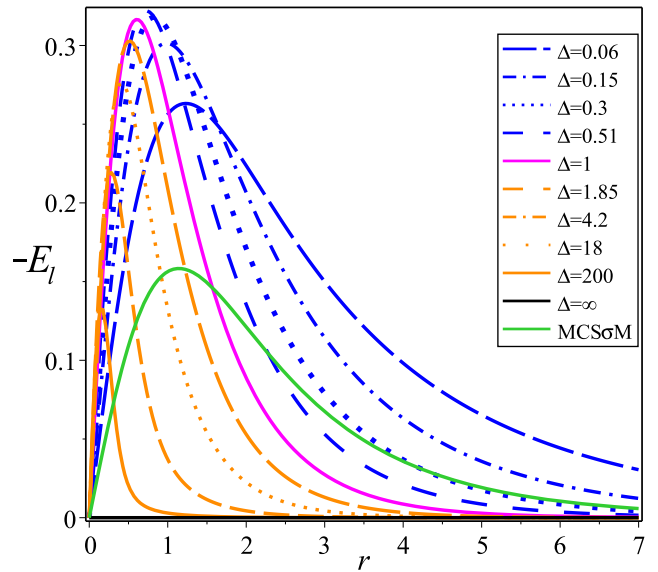}
\caption{Electric field $E_{{_l}}(r)=-A^{\prime }_0(r)$ profiles. }
\label{El_BPS}
\end{figure}

Figure \ref{El_BPS} describes the behavior for the electric field. For $n=1$%
, the maximum electric field amplitude is reached for some value $%
\Delta^{\ast }$ such that $0.5<\Delta ^{\ast }<1$. For $0<\Delta <1$, the
profiles become radially more spread out for decreasing $\Delta$ values,
i.e., $\Delta \rightarrow 0$. On the other hand, for $\Delta >1,$ the
profiles are located closer to the origin, being narrower, with their
amplitude decaying rapidly for increasing values of $\Delta $. In the limit $%
\Delta \rightarrow \infty$, the electric field disappears, which agrees with
electrically uncharged vortices.

\begin{figure}[]
\centering\includegraphics[width=8.5cm]{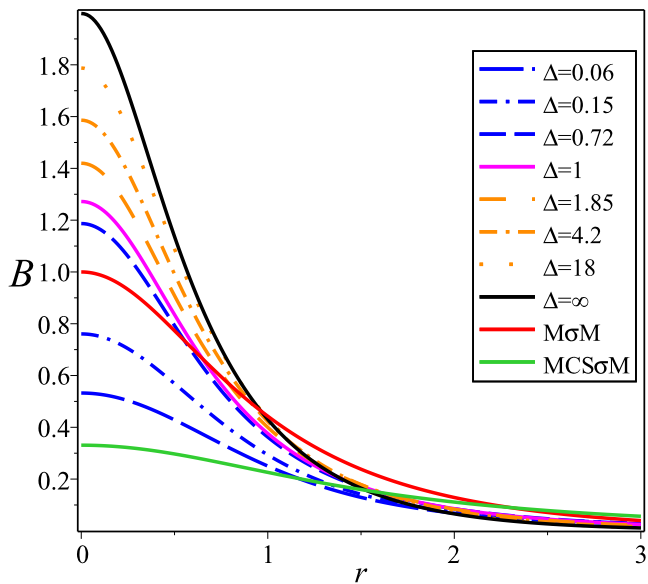}
\caption{Magnetic field ${B}(r)$ profiles.}
\label{B_BPS}
\end{figure}

Figure \ref{B_BPS} shows the profiles for the magnetic field, which are
lumps centered at the origin for $n=1$. For $0<\Delta <1$ (blue lines), the
profiles are more spread out and their amplitude at the origin decreases
continuously when $\Delta $ goes to zero. For $\Delta >1$ (orange lines),
the profiles become narrower and attain higher amplitudes for progressively
increasing $\Delta $. Nevertheless, the maximum narrowness and amplitude are
reached in the limit $\Delta \rightarrow \infty $ (solid black line).
Similarly, as it occurs with the sigma and vector fields, the magnetic field
profiles are located between the models defined by $\Delta =1$ and $\Delta
\rightarrow \infty $. The magnetic field at the origin, $B(0)$%
 can be increased or reduced in relation to the Lorentz symmetric
case.

\begin{figure}[]
\centering
\includegraphics[width=8.5cm]{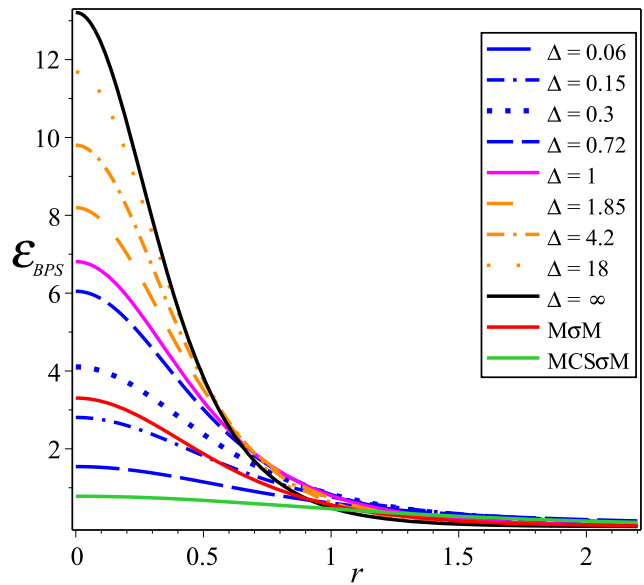}
\caption{Energy density $\protect\varepsilon_{_{BPS}}(r)$ profiles.}
\label{Energy_BPS}
\end{figure}

For $n=1$, the profiles of the BPS energy density (see Fig. \ref{Energy_BPS}%
) are lumps centered at the origin, such as the ones of the magnetic field.
For increasing values of $\Delta $, the amplitude grows at the
origin, the vortex core becomes more localized than the one of the
Lorentz symmetric counterpart (solid green or red lines). Similarly to the
magnetic field case, the maximum amplitude and narrowness occur in the limit
$\Delta \rightarrow \infty $.

By analyzing the profiles of the magnetic field and BPS energy density for $%
n>1$ and $\Delta $ finite, when the rotational symmetry is considered, a
ringlike profile is set out, whose values at origin and maximum amplitude
increase with $\Delta $. The ringlike structure of the magnetic field
mimics the behavior of the ones in models endowed with the Chern-Simons term
in the gauge sector, modifying the value and behavior at and near
the origin, however.

\section{Remarks and conclusions}

We have developed a comprehensive study about the electrically charged
self-dual configurations of the gauged $O(3)$ nonlinear sigma model
supplemented with a $CPT$-odd LV term in the gauge sector and a $CPT$-even
LV term in the sigma sector. We have verified that, for supporting charged
BPS solutions, the original model must be modified by introducing a neutral
scalar field with appropriate dynamics, in the same manner as it happens
with the Maxwell-Chern-Simons-Higgs model or the Maxwell-Chern-Simons $O(3)$
sigma model. We have managed to implement the Bogomol'nyi-Prasad-Sommerfield
formalism, finding the first order differential equations describing
self-dual charged configurations whose total energy is proportional to the
topological charge of the model, which gains LV contributions belonging to
the sigma sector. We have also observed that the total electric charge and
the total magnetic flux are related to each other such as it is shown in Eq.
(\ref{Rel}). These charged BPS configurations can be considered as classical
solutions related to an extended supersymmetric theory \cite{WittenOlive} in
a Lorentz-violating framework.

In particular, we have made an analysis of the axially symmetric vortex
solutions of the self-dual equations, demonstrating that the total BPS
energy, the magnetic flux and the electric charge are quantized
(proportional to the winding number) and also proportional to the LV
coefficients introduced in the sigma sector. A remarkable feature is that,
choosing some limits for the LV parameters, it is possible to
reproduce other gauged sigma models in the presence of Lorentz
violation, like the Maxwell-Chern-Simons $O(3)$ sigma model ($\Delta =1$)
and the gauged $O(3)$ sigma model ($\Delta \rightarrow \infty $) or
Chern-Simons $O(3)$ sigma model [for very large values of $(k_{AF})_{3}]$,
all them modified by Lorentz violation only in the sigma sector.
In general, Lorentz violation engenders altered solutions in
relation to the MCS$\sigma $M or M$\sigma $M %
profiles, as explicitly depicted in Figs. \ref{S_BPSx}--\ref{Energy_BPS}.
More specifically, LV affects the behavior of the magnetic field and BPS
energy density at the origin and near the origin : the amplitude
augments with $\Delta $, reaching its maximum deviation in the limit $\Delta
\rightarrow \infty $ , while the width decreases, yielding more compact and
localized vortex profiles (for large values of $\Delta )$. Thus, the
LV defects {have amplitude more} pronounced near the origin and are much
more localized than the Lorentz invariant solutions. An investigation to
be yet done concerns the influence of Lorentz symmetry violation on the
dynamics of these types of vortices.

\begin{acknowledgments}
R. C. and M. M. F. J. thank CNPq and FAPEMA (Brazilian agencies) for financial
support. C. F. F. and G. L. thank CAPES (Brazilian agency) for full financial support
\end{acknowledgments}


\begin{thebibliography}{99}
\bibitem{Gell-Mann} M. Gell-Mann and M. Levy, Nuovo Cimento \textbf{16}, 705 (1960)

\bibitem{Schwinger} J. Schwinger, Ann. Phys. (N.Y.) \textbf{2}, 407 (1957).

\bibitem{Polkinghorne} J. C. Polkinghorne Nuovo Cimento \textbf{8}, 179 (1958); \textbf{8}, 781 (1958).

\bibitem{polyakov} A. A. Belavin and A. M. Polyakov, JETP Lett. \textbf{22},
245 (1975).

\bibitem{CMP} R. Rajaraman, \textit{Solitons and Instantons}, (North-Holland,
Amsterdam, 1982); W. J. Zakrzewski, \textit{Low Dimensional Sigma Models} (Hilger,
Bristol, 1989).

\bibitem{BPS} E. B. Bogomol'nyi, Sov. J. Nucl. Phys. \textbf{24}, 449 (1976);
M. Prasad and C. Sommerfield, Phys. Rev. Lett. \textbf{35}, 760 (1975).

\bibitem{zakrzewski} R. A. Leese, M. Peyrard, and W. J. Zakrzewski,
Nonlinearity \textbf{3}, 387 (1990).

\bibitem{schroers} B. J. Schroers, Phys. Lett. B \textbf{356}, 291 (1995).

\bibitem{ghosh} P. K. Ghosh and S. K. Ghosh, Phys. Lett. B \textbf{366}, 199
(1996).

\bibitem{mukherjee1} P. Mukherjee, Phys. Lett. B \textbf{403}, 70 (1997).

\bibitem{mukherjee2} P. Mukherjee, Phys. Rev. D \textbf{58}, 105025 (1998).

\bibitem{sigmaMCSH1} K. Kimm, K. Lee, and T. Lee, Phys. Rev. D \textbf{53}, 4436
(1996).

\bibitem{sigmaMCSH2} J. Han and H.-S. Nam, Lett. Math. Phys. \textbf{73}, 17 (2005).

\bibitem{almeida} F. S. A. Cavalcante, M.S. Cunha, and C.A.S. Almeida, Phys.
Lett. B \textbf{475}, 315 (2000); M. S. Cunha, R. R. Landim, and C. A. S. Almeida,
Phys. Rev. D \textbf{74}, 067701 (2006).

\bibitem{kost97} D. Colladay and V. A. Kostelecky, Phys. Rev. D \textbf{55},
6760 (1997); \textbf{58},
116002 (1998).

\bibitem{coleman} S. R. Coleman and S. L. Glashow, Phys. Rev. D \textbf{59},
116008 (1999).

\bibitem{LV1} F. R. Klinkhamer and M. Schreck, Nucl. Phys. B \textbf{848}, 90 (2011);
M. Schreck, Phys. Rev. D \textbf{86}, 065038 (2012); M.A. Hohensee, R. Lehnert, D. F.
Phillips, and R. L. Walsworth, Phys. Rev. D \textbf{80}, 036010 (2009); A. Moyotl, H.
Novales-S\'{a}nchez, J. J. Toscano, and E. S. Tututi, Int. J. Mod. Phys. A
\textbf{29}, 1450039 (2014); \textbf{29}, 1450107 (2014); M. Cambiaso, R. Lehnert, and R.
Potting, Phys. Rev. D \textbf{90}, 065003 (2014); R. Bufalo, Int. J. Mod. Phys. A \textbf{29},
1450112 (2014); C.M. Reyes, L. F. Urrutia, and J. D. Vergara, Phys. Rev. D
\textbf{78}, 125011 (2008); Phys. Lett. B \textbf{675}, 336 (2009); C. M. Reyes, Phys. Rev. D
\textbf{82}, 125036 (2010); \textbf{80}, 105008 (2009); \textbf{87}, 125028 (2013).

\bibitem{LV2} V. A. Kostelecky and C. D. Lane, J. Math. Phys. (N.Y.) \textbf{40},
6245 (1999); R. Lehnert, J. Math. Phys. (N.Y.) \textbf{45}, 3399 (2004); D. Colladay
and V. A. Kostelecky, Phys. Lett. B \textbf{511}, 209 (2001); V. A. Kostelecky, C. D.
Lane, and A. G. M. Pickering, Phys. Rev. D \textbf{65}, 056006 (2002); C. D. Carone,
M. Sher, and M. Vanderhaeghen, Phys. Rev. D \textbf{74}, 077901 (2006); W.F. Chen and
G. Kunstatter, Phys. Rev. D \textbf{62}, 105029 (2000); O. M. Del Cima, D. H. T.
Franco, A. H. Gomes, J. M. Fonseca, and O. Piguet, Phys. Rev. D \textbf{85}, 065023
(2012); T. R. S. Santos and R. F. Sobreiro, Phys. Rev. D \textbf{91}, 025008 (2015).

\bibitem{LV3} M. A. Anacleto, F. A. Brito, and E. Passos, Phys. Rev. D \textbf{86},
125015 (2012); M. A. Anacleto, Phys. Rev. D \textbf{92}, 085035 (2015); E. O. Silva
and F. M. Andrade, Europhys. Lett. \textbf{101}, 51005 (2013); F.M. Andrade, E. O.
Silva, T. Prud\^{e}ncio, and C. Filgueiras, J. Phys. G \textbf{40} 075007 (2013).

\bibitem{scalar} M. N. Barreto, D. Bazeia, and R. Menezes, Phys. Rev. D \textbf{%
73}, 065015 (2006); A. de Souza Dutra, M. Hott, and F. A.Barone, Phys. Rev.
D \textbf{74}, 085030 (2006); A. de Souza Dutra and R. A. C. Correa, Phys. Rev.
D \textbf{83}, 105007 (2011); R. A. C. Correa, R. da Rocha, and A. de Souza
Dutra, Ann. Phys. (Amsterdam) \textbf{359}, 198 (2015).

\bibitem{seifert} M.D. Seifert, Phys. Rev. Lett. \textbf{105}, 201601
(2010); Phys. Rev. D \textbf{82}, 125015 (2010).

\bibitem{barraz} N.M. Barraz Jr., J.M. Fonseca, W.A. Moura-Melo, and J. A.
Helayel-Neto, Phys.Rev. D \textbf{76}, 027701 (2007); A. P. Baeta Scarpelli
and J. A. Helayel-Neto, Phys.Rev. D \textbf{73}, 105020 (2006).

\bibitem{oscillons} A. de Souza Dutra and R. A. C. Correa, Adv. High Energy
Phys. \textbf{2015}, 673716 (2015); R. A. C. Correa, R. da Rocha, and A. de Souza Dutra,
Phys. Rev. D \textbf{91}, 125021 (2015).

\bibitem{miller} C. Miller, R. Casana, M. M. Ferreira Jr., and E. da Hora,
Phys. Rev. D \textbf{86}, 065011 (2012).

\bibitem{casana1} R. Casana, M. Ferreira Jr., E. da Hora, and C. Miller,
Phys. Lett. B \textbf{718}, 620 (2012).

\bibitem{sourrouille} L. Sourrouille, Phys. Rev. D \textbf{89}, 087702
(2014); R. Casana and L. Sourrouille, Phys. Lett. B \textbf{726}, 488 (2013).

\bibitem{hott} C.H. Coronado Villalobos, J.M. Hoff da Silva, M.B. Hott, and H.
Belich, Eur. Phys. J. C \textbf{74}, 27991 (2014).

\bibitem{belich} H. Belich, F.J.L. Leal, H.L.C. Louzada, and M.T.D. Orlando,
Phys. Rev. D \textbf{86}, 125037 (2012).

\bibitem{Guillermo} R. Casana and G. Lazar, Phys. Rev. D \textbf{90}, 065007
(2014).

\bibitem{Claudio} R. Casana, C. F. Farias, and M. M. Ferreira Jr., Phys.
Rev. D \textbf{92}, 125024 (2015).

\bibitem{Jackiw} S. M. Carroll, G.B. Field, and R. Jackiw, Phys. Rev. D
\textbf{41}, 1231 (1990).

\bibitem{lee} C. K. Lee, K.M. Lee, and H. Min, Phys. Lett. B \textbf{252}, 79
(1990).

\bibitem{bolog} S. Bolognesi and S.B. Gudnason, Nucl. Phys. B \textbf{805}, 104
(2008).

\bibitem{ANO} A. A. Abrikosov, Zh. Eksp. Teor. Fiz. \textbf{32}, 1442
(1957) [Sov. Phys. JETP \textbf{5}, 1174 (1957)]; H. Nielsen and P. Olesen, Nucl.
Phys. \textbf{B61}, 45 (1973).

\bibitem{WittenOlive} E. Witten and D. Olive, Phys. Lett. B \textbf{78}, 97
(1978).

\end{thebibliography}
\end{document}